\documentclass[prl,twocolumn,showpacs]{revtex4}
\usepackage{graphicx,epsfig}
\usepackage{times,amsmath,amssymb,latexsym}
\usepackage{rotating}
\usepackage{stmaryrd}

\newcommand{\eqn}[1]{\begin{eqnarray} #1 \end{eqnarray}}
\newcommand{\abs}[1]{\left| #1 \right|}

\begin{document}

\title{Strongly Correlated States of Ultracold Rotating Dipolar
Fermi Gases}
\author{Klaus Osterloh$^{1}$, Nuria Barber\'{a}n$^{2}$, and Maciej Lewenstein$^{3}$}
\affiliation{(1) Institute for Theoretical Physics, University of Hannover, Appelstr. 2,
D-30167 Hannover, Germany\\
(2) Dept.~ECM, Facultat de F\'{i}sica, Universitat de Barcelona, E-08028
Barcelona, Spain and\\
(3) ICFO - Institut de Ci\`{e}ncies Fot\`{o}niques, Parc Mediterani de la Tecnologia,  Spain}

\begin{abstract}
We study strongly correlated ground and excited states of rotating quasi-2D Fermi gases
constituted of a small number of dipole-dipole interacting particles with dipole moments polarized
perpendicular to the plane of motion. As the number of atoms grows, the system enters {\it an intermediate regime}, where
ground states are subject
to a competition between distinct bulk-edge configurations. This effect obscures their description
in terms of composite fermions and leads to the appearance of novel composite fermion quasi-hole states.
In the presence of dipolar
interactions, the principal Laughlin state at filling $\nu=1/3$ exhibits a substantial energy gap for neutral
(total angular momentum conserving) excitations,
and is well-described as an incompressible Fermi liquid. Instead, at lower fillings, the ground state structure
favors crystalline order.
\end{abstract}

\pacs{03.75.Ss, 73.43.-f}
\date{\today }
\maketitle

%\email[E-mail:~]{Klaus.Osterloh@itp.uni-hannover.de}

%\email[E-mail:~]{Maciej.Lewenstein@itp.uni-hannover.de}

%\preprint{}
%\preprint{}

%Introduction\ldots

Some of the most fascinating challenges of modern atomic and molecular physics
arguably concern ultracold dipolar quantum gases \cite{Baranov:2002}. The recent experimental realization of a quantum degenerate dipolar Bose
gas of Chromium \cite{Griesmaier:2005}, and the progress in trapping and cooling of dipolar molecules \cite{SpecialIssueEurphysD04}
have opened the path towards ultracold quantum gases with dominant dipole interactions. Particularly interesting in this context
are {\em rotating dipolar gases} (RDG). Bose-Einstein condensates of RDGs exhibit novel forms of vortex lattices, e.~g., square, stripe- and
bubble-``crystal'' lattices  \cite{Cooper:2005}. The stability of these phases in the lowest Landau level was recently investigated
\cite{Komineas:2006}. We have demonstrated that the quasi-hole gap survives the large $N$ limit
for fermionic RDGs \cite{Baranov:2005a}. This property makes them perfect candidates to approach the strongly correlated regime,
and to realize Laughlin liquids (cf. \cite{Girvin}) at filling $\nu=1/3$,
and quantum Wigner crystals at $\nu\le 1/7$ \cite{Fehrmann:2006aa} for a mesoscopic number of atoms
$N\simeq 50-100$. Lately, Rezayi et al.~\cite{Rezayi:2005} have shown that
the presence of a small amount of dipole-dipole interactions stabilizes the
so-called bosonic Rezayi-Read state at $\nu=3/2$ whose excitations are
both fractional and non-Abelian.

%%%%%%%%%%%%%%%%%%%%%%%%%%%%%%%%%%%%%%%%%%%%%%%%%%%%%%%%%%%%%%%%%%%%%%%%%%%%%%%%
%
% What we do in this letter
%
In this letter, we investigate quasi-2D microscopic and mesoscopic clouds of
fermions whose dipole moments are polarized
perpendicular to their plane of motion.
Results are obtained by exact diagonalization of the Hamiltonian of the system for $N=3-12$ particles.
We study the lowest Landau level (LLL) physics, in particular the crossover from the weakly interacting Fermi liquid regime
to strongly correlated fractional quantum Hall-like states.
As $N$ grows, the ground state particles either organize themselves in the bulk or crystallize on the edge. This
competition is based on inter-particle correlations and exceeds the scope of the effective theory of composite fermions \cite{Jain}.
It leads to the appearance of novel ground states with a quasi-hole at the origin which are linked to unfavored composite-fermion states.
The analysis of the principal Laughlin state at filling $\nu=1/3$ reveals that
its low energy excitations in the angular momentum subspaces where quasi-holes are to be found,
correspond to edge excitations rather than topological quasi-hole defects.
Instead, ``neutral'' particle-hole pair excitations are already substantially gapped in these microscopic systems.

%%%%%%%%%%%%%%%%%%%%%%%%%%%%%%%%%%%%%%%%%%%%%%%%%%%%%%%%%%%%%%%%%%%%%%%%%%%%%%%%
%
% The Rotating Ultra-Cold Fermi-Gas
%
We consider a system of $N$ dipolar fermions rotating in an axially
symmetric harmonic trapping potential strongly confined in the direction
of the axis of rotation. Along this $z$-axis, the dipole moments, as well as spins are
assumed to be aligned. Various ways of experimental realizations of
ultracold dipolar gases are discussed in \cite{Baranov:2002}. In case of
low temperature $T$ and weak chemical potential $\mu$ with respect to the
axial confinement $\omega_z$, the gas is
effectively two-dimensional, and the Hamiltonian of the system in the
rotating reference frame reads
\eqn{{\cal{H}}=
     \sum_{j=1}^{N}\frac{1}{2M}
     \left(
     \vec{p}_j-M\Omega\vec{\mathbf{e}}_z\times\vec{r}_j
     \right)^2\!
     +\frac{M}{2}\left({\omega}_0^2-{\Omega}^2\right)r_j^2
     +V_d
     \label{Hamiltonian}}
Here, $\omega_{0}\ll \omega_{z}$ is the radial trap frequency, $\Omega$ is the
frequency of rotation, $M$ is the mass of the particles,
$V_d=\sum_{j<k}^N\frac{d^2}{\abs{\vec{r}_j-\vec{r}_k}^3}$ is the
dipolar interaction potential (rotationally invariant with respect to the $z$-axis), $d$ is the dipole moment, and $\mathbf{r}_{j}=x_{j}
\mathbf{e}_{x}+y_{j}\mathbf{e}_{y}$ is the position vector
of the $j$-th particle.
The first term of \eqref{Hamiltonian} is formally equivalent to the
{Landau} Hamiltonian of particles with mass
$M$ and charge $e$ moving in a
constant magnetic field of strength $B=2M\Omega c/e$
perpendicular to their plane of motion.
The spectrum of ${\mathcal{H}}_{\mathrm{Landau}}$
consists of equidistantly spaced and highly degenerate {Landau} levels (LL) with
energies $\varepsilon_{n}=\hbar\omega_c(n+1/2)$ where  $\omega_c=2\Omega$. We denote by
$N_{\mathrm{LL}}=1/2\pi l^{2}$ the number of states per unit area in each LL, where
$l=\sqrt{\hbar /m\omega_c}$ is the magnetic length. Given a homogeneous
fermionic surface density $n_f$, the filling factor $\nu
=2\pi l^{2}n_f$ can be defined and denotes the fraction of occupied Landau levels.
Even though the above definition applies to infinite homogeneous systems, it may be used for finite systems as a suitable
truncation of the Hilbert space at specific angular momenta.
The second term in \eqref{Hamiltonian} accounts for a rotationally induced
effective reduction of the trap strength. For critical rotation
$\Omega \rightarrow \omega_{0}$, the confining potential vanishes.

\begin{figure}[h]
\ \\[-3ex]
\centering
\caption{Interaction contribution to the GS energy as a function of $L^z$}
\includegraphics[width=0.45\textwidth,clip]{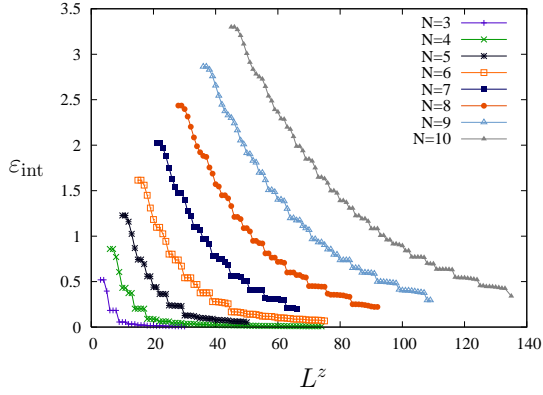}
%\ \\[-5ex]
%\includegraphics[width=0.16\textwidth,angle=-90]{FermiYrast.ps}
%\includegraphics[width=0.16\textwidth,angle=-90]{FermiGSL.ps}
\ \\[-3ex]
\label{Yrast}
\end{figure}

In the following, it is assumed that particles solely occupy the LLL \cite{valid}.
Restricted to the LLL and rewritten in second quantized form,
the Hamiltonian \eqref{Hamiltonian} reads
\eqn{\hat{\cal{H}}=\hbar\omega_0\hat{N}+\hbar\left(\omega_0-\Omega\right)\hat{L^z}
            +\!\!\!\!\!\!\sum_{m_1,\ldots,m_4}\!\!\!\!\!V_{1234}\,
             {\rm a}^\dagger_{m_1}{\rm a}^\dagger_{m_2}{\rm a}_{m_4}{\rm a}_{m_3}
             \label{Ham2nd}}
where $\hat{N}$ and $\hat{L}^z$ are the total number and z-component angular momentum
operators, $a^\dagger_{m_i}$ creates a particle with angular
momentum $m_i$,
and $V_{1234}=\frac{1}{2}\langle m_1\,m_2|\,V_{\rm d}\,|m_3\,m_4\rangle$
is the matrix element of interaction expressed in the Fock--Darwin single
particle angular momentum basis\cite{Fock}.
Due to circular symmetry, \eqref{Ham2nd}
can be diagonalized blockwise for a given $L^z$.
Calculations have been performed for $N=3$ to 12 particles with complete
and Davidson block diagonalization techniques \cite{dav}, respectively.
%%%%%%%%%%%%%%%%%%%%%%%%%%%%%%%%%%%%%%%%%%%%%%%%%%%%%%%%%%%%%%%%%%%%%%%%%%%%%%%%
%
% Lowest Landau level analysis, Yrast line and GSL crossover
%
In the LLL Hamiltonian, besides the first  constant term, there are two competing contributions to the energy:
the kinetic term linear in $\hat{L}$ and the
strength of interaction, which scales as
$d^2/l^3=2\hbar\omega_0(a_d/l)$ with $a_d=Md^2/\hbar^2$.
The natural unit of energy is $\hbar\omega_0$, whereas distances are measured
in $l$; from now on, we set $a_d/l\equiv 1$.
We analyze the ground state interaction
energy for a given $L^z$.
It reveals plateaus, clearly visible for small $N$(see Fig.~\ref{Yrast}).

\begin{figure}[h]
\ \\[-3ex]
\centering
\caption{Ground state
angular momentum series over $\alpha\equiv\omega_0-\Omega$ (the divergence at $\alpha=0$ is not shown).}
\includegraphics[width=0.45\textwidth,clip]{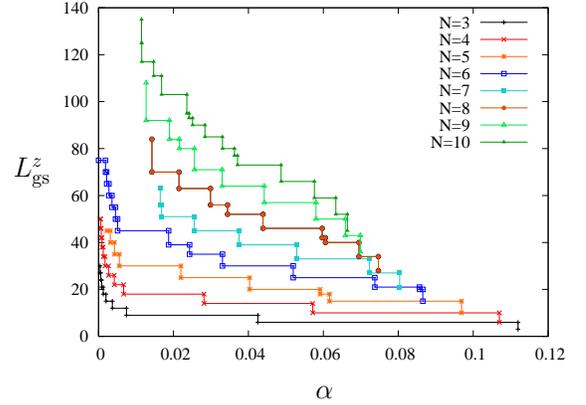}
\ \\[-3ex]
\label{GSL}
\end{figure}

Ground state candidates are the first states of these
plateaus where a downward cusp occurs in the spectrum.
By tuning the rotational frequency,
some of these states are selected as true ground states at specific
``magic'' angular momenta as depicted in Fig.~\ref{GSL}.
For relatively low values of $\Omega$, the ground state is the
filled LL state at $\nu=1$, which is completely insensitive to
type and strength of interaction as long as the LLL approximation holds.

When $\Omega$ is continuously increased, the system
evolves from the weakly interacting regime (where the second term of Eq.\eqref{Ham2nd} is larger than the third)
to strongly correlated states (where the interaction is  dominant).
In case of short-range interactions,
this process terminates at $L_{0}^z=\nu_0^{-1}N(N-1)/2$, where the fermionic (bosonic)
Laughlin states at filling $\nu_0=1/3(1/2)$ become the true ground states.

\begin{figure}[h]
\ \\[-0.5ex]
\centering
\caption{Ground state density-density correlation functions
$\hat{\rho}(\vec{r},\,\vec{r}_0)$ for $N=10$ dipolar fermions
at $L^z$=(top) 45,80, (centered) 90,93, (bottom)103,117 with $\vec{r}_0$ set to the
maximum of the density, which occurs at the edge.}
\includegraphics[width=0.49\textwidth,clip]{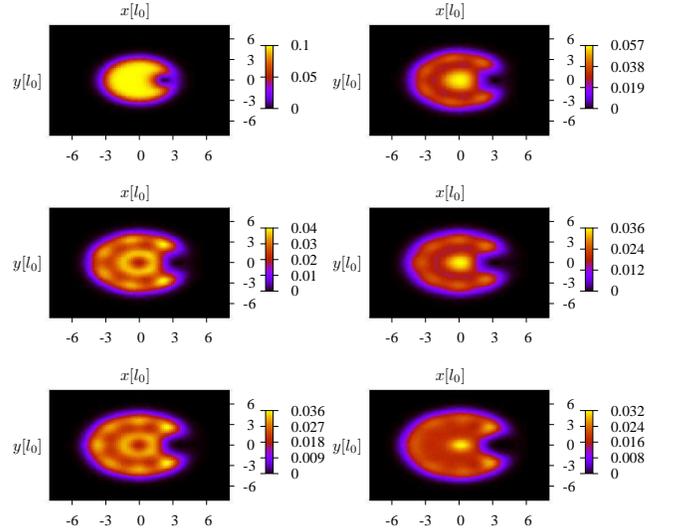}
\label{fermicorr}
\ \\[-1.5ex]
\end{figure}

The existence of a final $L^z$ is due to the fact that
the ground state contact interaction energy vanishes for $L^z\geq L_{0}^z$.
The long range nature of dipolar interactions lifts this
degeneracy. Thus, the whole principal series of fillings, i.e., $\nu=1/(2m+1)$
for fermionic gases, is accessible.
%%%%%%%%%%%%%%%%%%%%%%%%%%%%%%%%%%%%%%%%%%%%%%%%%%%%%%%%%%%%%%%%%%%%%%%%%%%%%%%%
%
% Ground state structure, composite fermion approach, quasi-hole ansatz
%
To reveal the internal structures of relevant states,
we consider the density-density correlation function $\hat{\rho}(\vec{r},\,\vec{r}_0)$, which represents the
conditional probability to find one atom at
$\vec{r}$ when another is simultaneously located at $\vec{r}_0,$
\eqn{\hat{\rho}(\vec{r},\,\vec{r}_0)=\sum_{j<k}^N\delta(\vec{r_j}-\vec{r})
     \delta(\vec{r_k}-\vec{r}_0)\,.}
It is crucial to analyze second order correlations, since  the GS density
only reveals radially symmetric contributions. Fig.~\ref{fermicorr} depicts
$\hat{\rho}(\vec{r},\,\vec{r}_0)$ for a selection of ground states
with $N=10$ particles.
Generally, for small $N$, starting from the unstructured maximum-density-droplet at $L^z=N(N-1)/2$ (top left)
with $\nu=1$, a fraction of atoms starts to arrange itself in a ``crystal''  on the
edge leaving a residual ``Fermi sea'' at the center(top right), until the correlations are
homogeneously established at $L_{0}^z=3N(N-1)/2$. For larger $N$, however, the energetically favorable number of edge atoms in the crystal changes "irregularly" with $L^z$ (right column);
the system seems to be ``frustrated'' in this respect.
In this regime,  the bulk has to reorganize itself accordingly, and novel fascinating states
with a hole at the center of the bulk appear (centered and bottom left).

In order to understand the ``magic'' $L^z$ numbers, Jain {\it et al.} \cite{Jain}
have proposed to model the interacting system in terms of an effective
theory of non-interacting composite fermions (CF) for electrons
in a quantum dot.
In this ansatz, each fermion captures an even number of  quantum fluxes, and the
wave function reads
\eqn{\Psi_{\mathrm{CF}}(\left\{ z_{j}\right\}) ={\mathcal{N}}
     {\hat{\mathcal{P}}}_{\mathrm{LLL}}\left\{
     \prod_{j<k}^{N}(z_{j}-z_k)^{2m}\Psi_{\mathrm{Landau}}\right\}\,.\label{psicf}
}
Here, $m$ denotes the number of  flux pairs, $\Psi_{\mathrm{Landau}}$
is an $N$-particle eigenfunction of ${\mathcal{H}}_{\mathrm{Landau}}$,
and $\hat{\mathcal{P}}_{\mathrm{LLL}}$
is the LLL projector.
Motivated by the success of CFs for fermions, Cooper and Wilkin have nicely adapted
this scheme to bosonic contact interacting gases\cite{WilkinCF}. Following this idea, we compared the series of true GS for non-interacting composites and dipolar fermions. As long as there is no real bulk in the system,  i.~e., for $N<7$,
the ground state series nearly identically match with the predictions of the effective CF theory.
Furthermore,  overlap calculations for electrons and short-range interacting bosons
\cite{Jain,WilkinCF} suggest that similar results will hold for dipolar fermions.
For bigger $N$,
 however, important deviations from CF theory, in particular in the intermediate regime, occur (see, for instance, the CF and true "magic" numbers in the table \ref{tabla}). This deficiency of the CF theory,
already commented  in Ref.~\cite{WilkinCF},
is more clearly related to the ``frustration'' effect and reorganization of the bulk in the present case.
For $N=10$, ground states with a density defect at the center are found at $L^z=90, 95$ and  $103$. They either have no ($L^z=95$), or are hardly similar to the corresponding "magic" CF states.
Instead, the analysis for the accessible range of systems strongly suggests that these states originate from
``parent'' states which are boosted by $N$ quanta of angular momentum, e.~g., $L^z=80,85$ and $93$ for the above series. None of these parent states has its "magic" analogues in the CF model.
\begin{widetext}
\ \\[-5ex]
\eqn{
\begin{array}{c|c|c|c|c|c|c|c|c|c|c|c|c|c|c|c|c|c|c|c|c|c|c|c}
\mathrm{L^z\ ,CF}\:\:&\!45\!&\!\!&\!55\!&\!\!&\!63\!&\!\!&\!69\!&\!\!&\!77\!&\!\!&\!83\!&\!\!&\!
90\!&\!\!&\!\!&\!97\!&\!103\!&\!111\!&\!117\!&\!125\!&\!135\!\\ \hline
\mathrm{L^z\ ,true}\:\:&\!45\!&\!52\!&\!\!&\!59\!&\!\!&\!66\!&\!\!&\!73\!&\!77\!&\!80\!&\!\!&\!85\!&\!
90\!&\!93\!&\!95\!&\!\!&\!103\!&\!111\!&\!117\!&\!125\!&135
\label{tabla}\end{array}}
\end{widetext}
%\eqn{\mathrm{CF}&&55, 63, 69, 77, 83, 90, 97, 103, 111, 117,125\\
%     \mathrm{true}&&52,60,66,73,77,80,85,90,93,95,103,111,117\nonumber}

To a good approximation, the close connection between parent and boosted state (e.g. $L=93$ and $103$ respectively) can be understood as a quasi-hole excitation,
which is analytically represented by
\eqn{\Psi_{\mathrm{qh}}(\left\{ z_{j}\right\}) ={\mathcal{N}}\left\{
     \prod_{j=1}^{N}(z_{j}-\zeta_0)^{m}\Psi_{\mathrm{L^z}}\right\}\,,\label{psicfhole}
}
where $\zeta_0=0$, and $m=1$ for $\nu=1/3$.
Indeed, the above wave function proves to be a good approximation for the states with $L^z=L^z_{parent}+N$: even though the exact states do not reveal a
true topological defect at the origin due to finite size, the density at the origin scales as $1/N$ and decreases from 0.14 to
$\simeq 0.10$ as $N$ varies from 7 to 11, while overlaps with the exact ground states grow from 0.6 to 0.7 despite
the simultaneous significant increase of the relevant Hilbert space dimension.

\begin{figure}[h]
\ \\[-2.5ex]
\centering
\caption{Density-density correlation function
$\hat{\rho}(\vec{r},\,\vec{r}_0)$ of the Laughlin state  for $N=12$ dipolar fermions with $\vec{r}_0$ chosen at the
maximum of the density, which occurs at the edge.
%; (b) Interaction spectrum for $N=6$ particles for the full basis (red) and fixed filling $\nu=1/3$ (green);
%quasi-hole states with $\nu=1/3-1/N$ are depicted in blue.
}
\includegraphics[width=0.45\textwidth,clip]{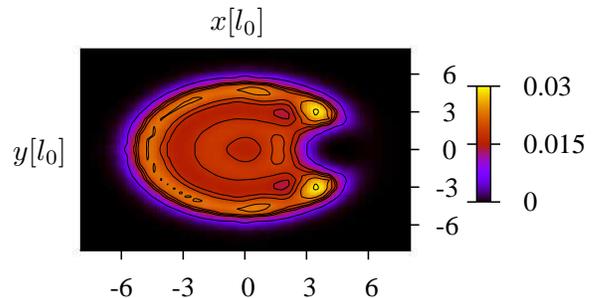}
\label{laughlin}
\ \\[-3.5ex]
\end{figure}

We have also studied in detail low energy excitations of the dipolar Laughlin state at $\nu=1/3$, which for $N\ge 10$  consists of
a significant bulk, surrounded by a practically ``melted crystal'' at the edge (see Fig.~\ref{laughlin}).
In order to relate our results to the ones obtained in the thermodynamical limit in Ref.~\cite{Baranov:2005a},
it is necessary to identify the finite-size quasi-hole excitations in the spectrum. To map out irrelevant candidates,
it is reasonable to appropriately truncate the Hilbert space by a finite size filling factor.
We use the approach of Ref.~\cite{Kasper}, and fix $\nu$ by imposing the constraint $m_{\nu}\le (N-1)/\nu $ on the maximum single particle angular momentum
in the $N$-particle Fock basis. If a quasi-hole (quasi-particle) is nucleated, the filling factor is lowered (raised) by a correction of the order of
$1/N$, i.~e., $m_{\nu}\rightarrow m_{\nu}\pm 1$. This significantly reduces the basis dimension of the system and maps out irrelevant states.
%We clearly observe the ``neutral'' gap and a tail of lower lying states at
%$L^z>L^z_{1/3}$.
If the rotational frequency $\Omega$ is tuned to favor $\Psi^L_{1/3}$ as the true ground state, the lower lying excitations
for $L^L_{1/3}<L^z<L^L_{1/3}+N$ turn out to be gapped by $\Delta \epsilon\simeq 1/N$.
For short-range interacting bosonic systems, these states have been partially identified as edge excitations that carry quanta of
angular momentum \cite{Cazalilla}.
To identify the excited states with a quasi-hole at the centre, the ground states at $L^z_{qh}=L^L_{1/3}+N$
are calculated with adapted $m_{\nu}+1$. These states closely follow the lowest energy branch of edge excitations,
and should be regarded as such, rather than quasi-holes. This is strongly supported by the fact that
the substantial density dip for $N=3$ continuously vanishes for increasing number of particles.
Additional information about these edge excitations can be obtained by direct "naked-eye" inspection of Fig. \ref{laughlin}. For the "standard" Laughlin state,  the edge excitations are described by a hydrodynamical  Luttinger liquid theory\cite{wens}, and their amplitude  decays as $\sin^{-3}(\phi)$, where $\phi$  is the "angular distance" from $\vec{r}_0$ along the edge; here the decay is much slower, which
is yet another signature of the finite size effects.

The above considerations imply that the only reliable quantity that remains to estimate the
quasi-hole energy in the large $N$ limit is the neutral gap at fixed $L^z$.  This gap remains more or less constant for smaller $N$s, and graduately increases as the number
of bulk particles in the system and  $N$ grow.  For the data of Ref.~\cite{Baranov:2005a} ($M=30$ a.m.u., $d=0.5$ Debye, and a trap frequency of $2\pi\times 10^{3}$ Hz),
the gap is of the order of percents of $2\hbar\omega_{0}$,
i.~e., it is substantial, but an order of magnitude less compared to the value estimated in \cite{Baranov:2005a} in the large $N$ limit.
Obviously, the lack of true bulk behavior in the investigated mesoscopic samples is responsible for this effect.

\begin{figure}[h]
\ \\[-2.5ex]
\centering
\caption{Radial density $\rho(r)$ of Laughlin states (circles) and true dipolar ground states (triangles) for $N=6$ at different filling factors.
Apart from five particles which constitute the edge ring, localization of the sixth particle at the origin is clearly visible for $\nu\leq 1/5$.}
\includegraphics[width=0.45\textwidth,clip]{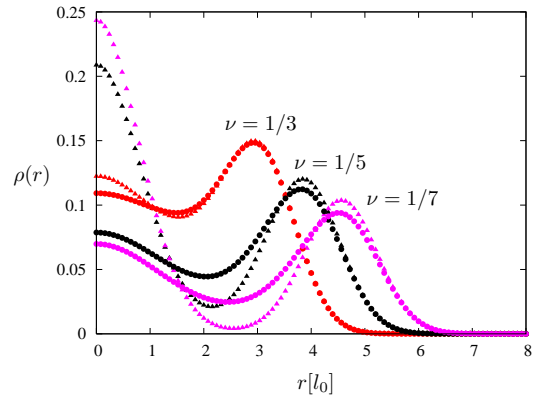}
\label{lowerfillings}
\ \\[-3.5ex]
\end{figure}

In the semiconductor fractional quantum Hall effect, Wigner crystals, i.~e., specific charge-density waves, were discussed as
competing ground states to Laughlin liquids. The interplay of quantum fluctuations with the interaction energy
have proven crystalline Wigner order to be favorable for low enough fillings for electrons.
This behavior intuitively changes in the case of dipolar particles as the interaction energy scales differently in the density of the particles.
A very recent detailed analysis of this issue has confirmed
the stability estimates for Laughlin and Wigner states of Ref.~\cite{Baranov:2005a}
for systems constituted of 50-200 particles \cite{Fehrmann:2006aa}.
Fehrmann et al.~ have proven that phonon excitations destabilize the Wigner crystal for fillings $\nu > 1/7$, which melts into a Laughlin liquid.
In the microsystems discussed in this letter, this crossover is strongly supported. For fillings $\nu\leq 1/5$, density profiles
significantly deviate from the Laughlin state and show a clear pattern of hexagonal order for $N=6$ particles
as depicted in Fig.~\ref{lowerfillings}
where the outer edge ring of the state is constituted of five particles. Even though speaking of a true ``crystal'' is certainly not applicable
in such a system, signatures of crystalline order are clearly present.

Summarizing, we have studied in detail ground and excited states of
quasi-2D ultracold rotating dipolar Fermi gases.
By exact diagonalization methods, we studied systems up to 12 particles. We have identified novel kinds of strongly correlated
states in the intermediate regime, i.~e., pseudo-hole excitations of CF states.
Calculation of the substantial gap in the excitation spectrum of the dipolar Laughlin state at $\nu=1/3$
prove the accessibility of fractional quantum Hall states in these microsystems.
At lower fillings, interactions favor crystalline order.
Rotating dipolar gases are thus very suitable candidates
to realize Laughlin-like and more exotic quantum liquids, as well as  their crossover behavior
to Wigner crystals.
\begin{acknowledgments}
We are indebted to W.~Apel for valuable advice and help, and we thank D. Dagnino,
M.~Leduc, A. Riera, C.~Salomon, L.~Sanchez-Palencia, and G.~V.~Shlyapnikov for helpful
discussions. We acknowledge support from the Deutsche Forschungsgemeinschaft (SPP1116 and SFB 407),
ESF PESC ``QUDEDIS'', EU IP ``SCALA'', Spanish MEC (FIS2004-05639, FIS2005-04627, 2005SGR00343
and Consolider Ingenio 2010 "QOIT").
\end{acknowledgments}

\end{document}